\documentclass[preprint]{elsarticle}
\usepackage{algorithm2e}
\usepackage[utf8]{inputenc}
\usepackage{graphicx
,amssymb,amsmath,tikz,tcolorbox,amsthm,lineno,url,fullpage}
\usepackage[]{algorithm2e}

\usepackage{url}

\newenvironment{claimproof}{\paragraph{Proof of the claim:}}{\hfill$\triangleleft$}

\date{}
\journal{}

\newtheorem{observation}{Observation}

\newtheorem{lemma}{Lemma}
\newtheorem{theorem}{Theorem}

\usepackage{thmtools}
\usepackage{thm-restate}

\usepackage{todonotes}

\usepackage{hyperref}
\usepackage{xcolor}
\hypersetup{
    colorlinks=true,
    linkcolor=blue,
    urlcolor=blue,
    citecolor=blue
}
\usepackage{cleveref}

\DeclareMathOperator{\MP}{mp}

\DeclareMathOperator{\MMP}{MP}

\begin{document}
\begin{frontmatter}

\title{Multipacking in Euclidean Metric Space\tnoteref{label1,label2}}
\tnotetext[label1]{A preliminary version of this paper entitled ``Multipacking in Euclidean Metric Space'' has appeared in the proceedings of the {\it Conference on Algorithms and Discrete Applied Mathematics} (CALDAM), pp 109–120, 2025 (\url{https://doi.org/10.1007/978-3-031-83438-7_10}). }

\author[1]{Arun Kumar Das}
\ead{arund426@gmail.com}

\author[2]{Sandip Das}
\ead{sandip.das.69@gmail.com}

\author[2]{Sk Samim Islam}
\ead{samimislam08@gmail.com}

\author[2]{Ritam Manna Mitra}
\ead{rmmitra98@gmail.com}

\author[3]{Bodhayan Roy}
\ead{bodhayan.roy@gmail.com}

\address[1]{University of Hyderabad, Hyderabad, India}
\address[2]{Indian Statistical Institute, Kolkata, India}
\address[3]{Indian Institute of Technology, Kharagpur, India}

\begin{abstract}
       
Here we study the  multipacking problems for geometric point sets with respect to their Euclidean distances. We consider a set of $n$ points $P$ and define $N_s[v]$ as the subset of $P$ that includes the $s$ nearest points of $v \in P$ and the point $v$ itself. We assume that the \emph{$s$-th neighbor} of each point is unique, for every $s \in \{0, 1, 2, \dots , n-1\}$.  
For a natural number $r \leq n-1$, an $r$-multipacking is a set $ M \subseteq P  $  such that for each point $ v \in P $ and for every integer $ 1\leq s \leq r  $, $|N_s[v]\cap M|\leq (s+1)/2$. The $r$-multipacking number of $ P $ is the maximum cardinality of an $r$-multipacking of $ P $ and is denoted by $ \MP_{r}(P) $. For $r=n-1$, an $r$-multipacking is called a multipacking and $r$-multipacking number is called as multipacking number.  For $r=1 \text{ and } 2$, we study the problem of computing a maximum $r$-multipacking of the point sets in $\mathbb{R}^2$. We show that a maximum $1$-multipacking can be computed in polynomial time but computing a maximum $2$-multipacking is \textsc{NP-hard}. Further, we provide approximation and parameterized solutions to the $2$-multipacking problem.


\end{abstract}

\end{frontmatter}

 \section{Introduction}
 \label{sec:introduction}
The problem of locating facilities, both desirable and undesirable, is of utmost importance due to its huge number of applications in the real world~\cite{wolf2011facility}. Allocation of restricted resources within an arena has been extensively studied across various models and methodologies since the 1980s. Drezner and Wesolowsky did a "maximin" problem where a certain facility must serve a given set of users but must be placed in such a manner that its minimum distance from the users is maximized (see~\cite{drezner1980maximin,drezner1983location}). Rodriguez et al.~\cite{rodriguez2006general} gave a generic model for the obnoxious facility location problem inside a given polygonal region. Rakas et al. \cite{rakas2004multi} devised a multi-objective model to determine the optimal locations for undesirable facilities.  The book by Daskin \cite{daskin1997network} contains a nice collection of problems, models, and algorithms regarding obnoxious facility location problems in the domain of supply chain networks. In this paper, we study the \emph{Multipacking} \cite{teshima2012broadcasts} problem in Euclidean space, which resembles the obnoxious facility location problem in computational geometry. 
 


Given a point set $P \subseteq \mathbb{R}^2$ of size $n$, the \emph{neighborhood of size $r$} of a point $v \in P$ is the subset of $P$ that consists of the $r$ nearest points of $v$ and $v$ itself. This subset is denoted by $N_r[v]$. A natural and realistic assumption here is the distances between all pairs of points are distinct. Thus the $r$-th neighbor of every point is unique. This assumption is easy to achieve through some form of controlled perturbation (see \cite{mehlhorn2006reliable}).
For a natural number $r \leq n-1$, an \textit{$r$-multipacking} is a subset $ M $ of $ P  $, such that for each point $ v \in P $ and for every integer $ 1\leq s \leq r  $, $|N_s[v]\cap M|\leq (s+1)/2$, i.e. $M$ contains at most half of the number of elements of $N_s[v]$.
 The \emph{$r$-multipacking number} of $ P $ is the maximum cardinality of an $r$-multipacking of $ P $ and is denoted by $ \MP_r(P) $.

An $(n-1)$-multipacking $M$ of $P$ is referred to as a \emph{multipacking} of $P$. The \emph{multipacking number} of $P$, denoted by $\MP(P)$, refers to the cardinality of a maximum multipacking of $P$. Computing maximum multipacking  involves selecting the maximum number of points from the given point set satisfying the constraints of multipacking.
This variant of facility location problem is interesting due to their wide application in operation research~\cite{rodriguez2006general}, resource allocation~\cite{melachrinoudis1999bicriteria}, VLSI design  \cite{abravaya2010maximizing}, network planning~\cite{cappanera2003discrete,rakas2004multi}, and clustering  \cite{henzinger2020dynamic}.

To the best of our knowledge, no significant work on multipacking has been done on planar point sets concerning their Euclidean distances. Whereas the literature contains numerous intriguing findings for this problem in graph theory. Multipacking was first discussed by Teshima~\cite{teshima2012broadcasts}. Other related works can be found in~\cite{beaudou2019multipacking,cornuejols2001combinatorial,das2023relation,das2024cactusmultipacking,meir1975relations}. However, as of now, no polynomial-time algorithm is known for finding a maximum multipacking of general graphs, the problem is also not known to be \textsc{NP-hard}. Nonetheless, polynomial-time algorithms exist for trees and strongly chordal graphs~\cite{brewster2019broadcast}. For further references on algorithmic results regarding multipacking problem on graphs, we refer to the paper~\cite{foucaud2021complexity}. A survey on the multipacking on graphs is given in the book \cite{haynes2021structures}. 



\vspace{0.25cm}
\noindent\textbf{Our Contribution:} First, we study the multipackings of the point sets on $\mathbb{R}^1$. We start by bounding the multipacking number of $P\subset \mathbb{R}^1$ and we have shown the tightness of the bounds by constructing examples.

\begin{restatable}{theorem}{thmmponeDtightlowerupperbound}
\label{thm:mp 1D tight lower bound n/3 upper bound n/2}
    Let $P $ be a point set on $\mathbb{R}^1$. Then $\lfloor\frac{n}{3}\rfloor \leq \MP(P)\leq \lfloor \frac{n}{2}\rfloor$. 
    Both the bounds are tight.
\end{restatable}

We also provide an algorithm to find the maximum multipacking of a point set on a line.

\begin{restatable}{theorem}{rkthmonedcorrectness}\label{thm:rk1D}
For any $1 \leq r \leq n-1$, a maximum $r$-multipacking for a given point set $P \subset \mathbb{R}^1$ with $|P|=n$ can be computed in $O(n^2r)$ time.
\end{restatable}

\noindent Theorem \ref{thm:rk1D} yields that, a maximum multipacking for a given point set $P \subset \mathbb{R}^1$ with $|P|=n$ can be computed in $O(n^3)$ time, since an $(n-1)$-multipacking is nothing but a multipacking.



Further, we define the function $\MMP_{r}(t)$ to be the smallest number such that any point set $P \subset \mathbb{R}^2$ of size $ \MMP_{r}(t)$ admits an $r$-multipacking of size at least $t$. We denote $\MMP_{n-1}(t)$ as $\MMP(t)$, i.e. $\MMP(t)$ is the smallest number such that any point set $P \subset \mathbb{R}^2$ of size $ \MMP(t)$ admits a \emph{multipacking} of size at least $t$. It can be followed that $\MMP(1)=1$ from the definition of multipacking. No general solution has been found for this problem yet. 


We provide the value of $\MMP(2)$ in the following theorem:

\begin{restatable}{theorem}{MPtwo}\label{thm:MP2}
    $\MMP(2)=6$.
\end{restatable}


Until now, there is no known polynomial-time algorithm to find a maximum multipacking for a given point set $P\subset \mathbb{R}^2$, and the problem is also not known to be \textsc{NP-hard}. In this paper we make an important step towards solving the problem by studying the $1$-multipacking and $2$-multipacking in $\mathbb{R}^2$. We considered the NNG (Nearest-Neighbor-Graph) $G$ of $P\subset \mathbb{R}^2$. Using the observation, the maximum independent sets of $G$ are the maximum $1$-multipackings of $P$, we conclude that a maximum $1$-multipacking can be computed in polynomial time. However the $2$-multipacking problem in $\mathbb{R}^2$ is much more complicated. We have studied the hardness of the $2$-multipacking problem in $\mathbb{R}^2$. We reduce the well-known \textsc{NP-complete} problem Planar Rectilinear Monotone 3-SAT~\cite{de2012optimal} to our problem to show the following. 

\begin{restatable}{theorem}{NPC}\label{thm:NPC}
Computing a maximum $2$-multipacking of a given set of $n$ points $P$ in $\mathbb{R}^2$ is \textsc{NP-hard}.
\end{restatable}



Moreover, we provided approximation and parameterized solutions to this problem. We have shown that a maximum $2$-multipacking of a given set of points $P$ in $\mathbb{R}^2$ can be approximated in polynomial time within a ratio arbitrarily close to $4$ and $2$-multipacking of size $k$ of a given set of $n$ points in $\mathbb{R}^2$ can be computed in time $18^kn^{O(1)}$, if it exists.




\vspace{0.25cm}
\noindent\textbf{Organisation:} In Section \ref{sec:Multipacking in geometry}, we provide an algorithm to find a maximum $r$-multipacking of a point set on a line and we prove some bounds on the multipacking number on a line. In Section \ref{sec:MP}, we prove that   $\MMP(2)=6$ using some geometric observations.  In Section \ref{sec:1,1/2 Multipacking in a 2D plane},  we have shown that a maximum $1$-multipacking in $\mathbb{R}^2$ can be computed in polynomial time. In Section \ref{sec:2,1/2 Multipacking in a 2D plane}, we study the hardness of the $2$-multipacking problem in $\mathbb{R}^2$. We prove that the problem is \textsc{NP-hard}. Moreover, we provide approximation and parameterized solutions to this problem. We conclude in Section \ref{sec:conclusion}.










\section{Maximum multipacking in $\mathbb{R}^1$}\label{sec:Multipacking in geometry}

    


\subsection{Bound of maximum multipacking in $\mathbb{R}^1$}\label{subsec:Bounds of Multipacking in R1}
We start by bounding the multipacking number of a point set in the following lemma.

\begin{lemma}
\label{lem:mp 1D lower bound n/3 upper bound n/2}
    Let $P $ be a point set on $\mathbb{R}^1$. Then $\lfloor\frac{n}{3}\rfloor\leq \MP(P)\leq \lfloor\frac{n}{2}\rfloor$. 
\end{lemma}

\begin{proof}
    In order to show the lower bound, we show that for any point set on $\mathbb{R}^1$ of size $n$ we can always find a multipacking of size $\lfloor\frac{n}{3}\rfloor$. Let $P=\{p_1, p_2, \dots, p_n\}$ be $n$ points on $\mathbb{R}^1$ sorted with respect to the ascending order of their $x$-coordinates. Construct a set $M$ as such, $M=\{p_1, p_4, p_7, \dots, p_{3k+1} \}$ (where $k \in \mathbb{N}$ and $3k+1 \leq n < 3(k+1) + 1$) is a multipacking of $P$. Note that $M$ is a multipacking since for any $1 \leq s \leq n-1$ and $p_i\in P$, $N_s[p_i]$ contains at most $\lfloor\frac{s+1}{2}\rfloor$ points from $M$. This shows the lower bound of the lemma.  The upper bound follows from the definition of multipacking. 
\end{proof}




\noindent \textbf{Tight bound example for the upper bound:} For the upper bound of Lemma \ref{lem:mp 1D lower bound n/3 upper bound n/2}, we consider the point set $P = \{ p_1, \ldots , p_n \}$ where, $$p_i=
    \begin{cases}
        0 & \text{if $i=0$ }\\
        \frac{4}{3}(2^{i-1}-1)-\frac{i-1}{2} & \text{if $i$ is odd} \\
        \frac{4}{3}(2^{i}-1)-\frac{i}{2} - 2^{i-1} +1 & \text{if $i$ is even, $i>0$}
    \end{cases}$$

Note that  $p_i$'s in $P$ are in ascending order. From definition of this example, $p_{2k+2}-p_{2k+1}>p_{2k+1}-p_{1}$ and $p_{2k+2}-p_{2k+1}>p_{2k+3}-p_{2k+2}$ for each $k$. From this, we conclude that the set $M = \{ p_1, p_4, p_6, p_8, \ldots \}$ is a maximum multipacking and $|M| = \lfloor\frac{n}{2}\rfloor$.


\vspace{0.25 cm}
\noindent \textbf{Tight bound example for the lower bound: }
For the lower bound of Lemma \ref{lem:mp 1D lower bound n/3 upper bound n/2}, we consider the point set $P = \{ p_1, \ldots , p_n \}$ where $p_i = 2^i$. 

Note that $p_{i-1}-p_1 < p_i - p_{i-1}$ holds for each $p_i$ in $P$. From this, we conclude that the set $M = \{p_1, p_4, p_7, \ldots \}$ is a maximum  multipacking with $|M| = \lfloor\frac{n}{3}\rfloor$.


Combining these arguments together with Lemma \ref{lem:mp 1D lower bound n/3 upper bound n/2} we have the following theorem.

\thmmponeDtightlowerupperbound*


\subsection{Algorithm to find a maximum multipacking in $\mathbb{R}^1$}\label{subsec:AlgoR1}

We devise algorithm \ref{algo:2_mp} that checks possible violations for every member of $P$ and selects the eligible candidates as multipacking accordingly. It takes a point set $P \subset \mathbb{R}^1$ sorted in ascending order with respect to their $x$-coordinates  as input and outputs a maximum multipacking $M$ where $M \subseteq P$.

\medskip
\RestyleAlgo{ruled}    
\begin{algorithm}[H]
    \SetAlgoLined
    \KwIn{A point set $P$ in $\mathbb{R}$ and $M \subseteq P$ and the value of $r$.}
    \KwOut{\textit{if} $M$ is a valid multipacking on $P$ \textit{then} return $1$,\ \textit{else} return $0$}
    \For{$p_i$}{
      \For{$s=1 \rightarrow r$}{
      \eIf{$|N_s[p_i] \cap M| \leq (s+1)/2$}{continue;}{return $0$;}
      }
       
    }
    return $1$\;

 \caption{Check $r$-multipacking - $CMP_{r}(P,M)$}
 \label{algo:chk_2_mp}
\end{algorithm}
\medskip

Algorithm \ref{algo:chk_2_mp} ($CMP_{r}(P,M)$) is designed to check the eligibility of a member of $P$ as a point in multipacking and executed as a subroutine in each iteration of \cref{algo:2_mp}. $CMP_{r}(P,M)$(Algorithm \ref{algo:chk_2_mp}) takes a point set $M \subseteq P$ as input and outputs $1$ if $M$ is a multipacking and $0$ if not. 
Given a point set $P = \{ p_1, p_2, \hdots , p_n \}$, at the $i^{\text{th}}$ iteration, $p_i$ is included in $M$ and $CMP_{r}(P,M)$ is called as a subroutine. Depending on the eligibility check done by this subroutine $M$ is formed. The algorithm terminates when $P$ is exhausted and returns $M$ as output. 

\medskip
\RestyleAlgo{ruled}    
\begin{algorithm}[H]
    \SetAlgoLined
    \KwIn{A point set $P$ in $\mathbb{R}$ and the value of $r$.}
    \KwOut{A maximum multipacking $M$ of $P$.}
    $M=\phi$\; $i=1$\;
    \For{$p_i$}{
        $M = M \cup \{p_i\}$\;
       \If{$CMP_{r}(P,M) ==0$}{$M = M \setminus \{p_i\}$\;}
       $i++$\;
    }
    return $M$\;
 \caption{Maximum $r$-multipacking}
 \label{algo:2_mp}
\end{algorithm}



\rkthmonedcorrectness*

\begin{proof}
 We prove the result by showing the correctness of \cref{algo:2_mp} that returns a maximum $r$-multipacking of the given point set $P$ as output in $O(n^2r)$ time. For simplicity, we just say multipacking in place of $r$-multipacking in this proof. Let $P=\{a_1,a_2,\dots,a_n\}$ where $a_{i}<a_{i+1}$ for each $i$. Let $M$ be a solution of \cref{algo:2_mp}. According to the algorithm, we can say that $M$ is a multipacking. Let us assume that $M$ is not a maximum multipacking. Let $M'$ denote a maximum multipacking of the point set $P$. To compare $M$ and $M'$, we consider the members of $M$ and $M'$ sorted in ascending order with respect to their $x$-coordinate. Let $M=\{a_{g_1},a_{g_2},\dots,a_{g_m}\}$ and $M'=\{a_{h_1},a_{h_2},\dots,a_{h_{m'}}\}$ where $a_{g_i}<a_{g_{i+1}}$ and $a_{h_j}<a_{h_{j+1}}$ for each $i$ and $j$. By assumption, $|M|<|M'|$ or $m<m'$. Let $t=\min\{i:a_{g_i}\neq a_{h_{i}}, 1\leq i\leq m\}$. If $a_{g_t}> a_{h_{t}}$, then $\{a_{h_1},a_{h_2},\dots,a_{h_{t}}\}$ is not a multipacking of $P$ since \cref{algo:2_mp} allocates multipacking for every possible point from left to right of $P$. Therefore, $a_{g_t}< a_{h_{t}}$. Let $M''=\{a_{g_1},a_{g_2},\dots,a_{g_t},a_{h_{t+1}},a_{h_{t+2}},\dots,a_{h_{m'}}\}$  (See Fig. \ref{algo:2_mp}). 

 \begin{figure}[htbp]
    \centering
    \includegraphics[width = .75\linewidth]{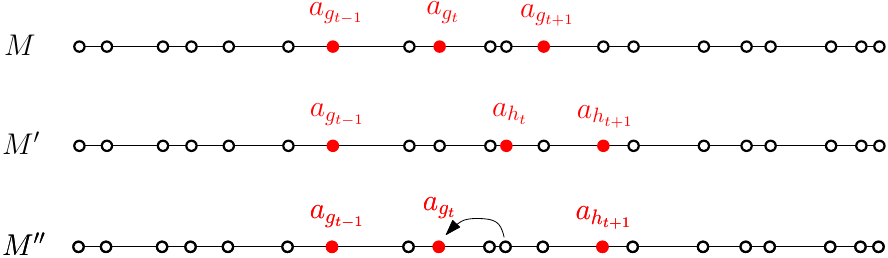}
    \caption{Correctness proof of algorithm \ref{algo:2_mp}.}
    \label{fig:MMM}
\end{figure}

 \vspace{0.2cm}
\noindent
\textbf{Claim \ref{thm:rk1D}.1. } $M''$ is a multipacking of $P$.

\begin{claimproof} We want to prove that $M''$ is a multipacking using the fact saying that $M$ and $M'$ are multipackings. Observe that, $N_s[a_i]$ is a list of consecutive points of $P$ for each $ a_i \in P $. Now fix a point $a_i\in P$. First consider $a_{g_t},a_{h_t}\in N_s[a_i]$ for some $s$. Then $|N_s[a_i]\cap M''|=|N_s[a_i]\cap M'|\leq (s+1)/2$. Suppose $a_{g_t}\in N_s[a_i]$ but $a_{h_t}\notin N_s[a_i]$. Then $N_s[a_i]\cap M''=N_s[a_i]\cap M$, this implies $|N_s[a_i]\cap M''|=|N_s[a_i]\cap M|\leq (s+1)/2$. If $a_{g_t}\notin N_s[a_i]$ but $a_{h_t}\in N_s[a_i]$, then $|N_s[a_i]\cap M''|<|N_s[a_i]\cap M'|\leq (s+1)/2$. Now we are left with only one case which says that $a_{g_t},a_{h_t}\notin N_s[a_i]$. In that case $N_s[a_i]\cap M''=N_s[a_i]\cap M'$, this implies $|N_s[a_i]\cap M''|=|N_s[a_i]\cap M'|\leq (s+1)/2$. Therefore, $|N_s[a_i]\cap M''|\leq (s+1)/2$ for every point $a_i\in P$ and for every integer $ 1\leq s \leq r$.   
\end{claimproof}

Since $|M''|=m'$, therefore $M''$ is a maximum multipacking of $P$. Similarly we can show that $\{a_{g_1},a_{g_2},\dots,$ $a_{g_t},a_{g_{t+1}},a_{h_{t+2}},\dots,a_{h_{m'}}\}$ is also a maximum multipacking using the fact: $M$ and $M''$ are multipackings. We can keep on doing the same method to arrive at the maximum multipacking $M_1=\{a_{g_1},a_{g_2},\dots,a_{g_{m-1}},a_{g_{m}},$ $a_{h_{m+1}},\dots,a_{h_{m'}}\}$. But \cref{algo:2_mp} ensures that one cannot add more points to the set $\{a_{g_1},a_{g_2},\dots,a_{g_{m-1}},a_{g_{m}}\}$ to construct a larger multipacking. Therefore, $|M_1|=|M|$, this implies $m=m'$ which is a contradiction. Therefore, \cref{algo:2_mp} returns a maximum $r$-multipacking.

    Now we analyze the running time of the \cref{algo:2_mp}. It is clear that the algorithm calls the subroutine $CMP_{r}(P,M)$ (\cref{algo:chk_2_mp}) for $n$ times. Further, the subroutine $CMP_{r}(P,M)$ (\cref{algo:chk_2_mp}) takes $O(nr)$ time to check the correctness of being multipacking for a set. Thus \cref{algo:2_mp} takes $O(n^2r)$ time to return a solution.  
 \end{proof}

It is worth mentioning that, computing the maximum multipacking set of a planar point set relates with the notion of independent set (which we show later in the paper) which becomes \textsc{NP-hard} in general. Thus it is important to study a bound for the multipacking number in $\mathbb{R}^2$. Next, we address the bound on the size of $P$ to achieve a multipacking of a desired size irrespective of the distribution of a set of points in $\mathbb{R}^2$.

\section{Minimum number of points to achieve a desired multipacking size in $\mathbb{R}^2$}
\label{sec:MP}

In this section, we study the $\MMP$ function. Recall the definition of the same. $\MMP_{r}(t)$ is the smallest number such that any point set $P \subset \mathbb{R}^2$ with $|P|= \MMP_{r}(t)$ admits an $r$-multipacking of size at least $t$. As stated earlier $\MMP_{n-1}(t)$ is denoted by $\MMP(t)$ and $\MMP(1)=1$. Now our goal is to find the value of $\MMP(2)$, which we state in the following theorem.




\begin{figure}[htbp]
\centering%
\begin{minipage}[t]{0.50\textwidth}
	\centering
	\includegraphics[width = .50\textwidth]{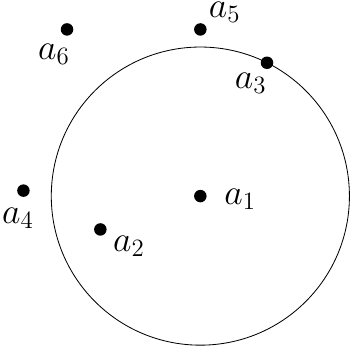}
	\caption{Any $6$ points always have $\MP(P) \geq 2$.}
	\label{fig:C_2[a_1]}
\end{minipage}%
\hfill%
\begin{minipage}[t]{0.45\textwidth}
	\centering
	\includegraphics[width = .65\textwidth]{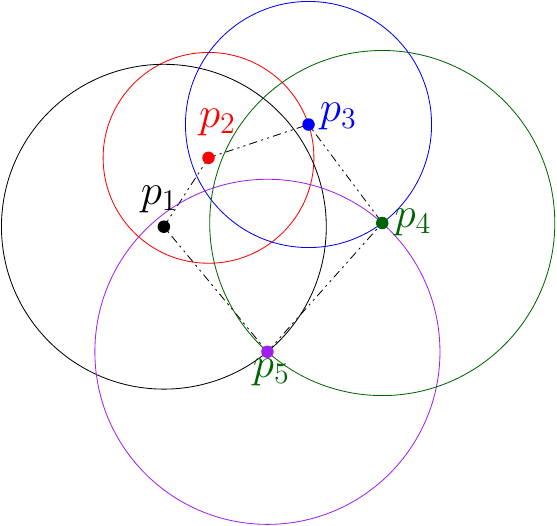}
	\caption{$5$ points with $\MP(P) =1$.}
	\label{fig:mp2}
\end{minipage}%
\end{figure}

\MPtwo*

\begin{proof}
    We prove this using contradiction. Suppose $\MMP(2)>6$. That means there is a point set $P=\{a_1,a_2,\dots,a_6\}\subset \mathbb{R}^2$ such that $\MP(P)=1$. This implies, for each $i,j(i\neq j)$ there exists $k$ such that $\{a_i, a_j\}\subset N_2[a_k]$, otherwise $\{a_i, a_j\}$ can form a multipacking of size $2$ which is not allowed. If $a_i$ and $a_j$ are the $1^{\text{st}}$ and $2^{\text{nd}}$ neighbor of $a_k$ respectively, we denote $C_2[a_k]$ to be the circle with center at $a_k$ passing through $a_j$. Without loss of generality we can assume that $C_2[a_1]$ is the circle with the largest radius among all the circles $C_2[a_k]$, $k\in \{1,2,\dots,6\}$ and moreover $a_2$ and $a_3$ are the $1^{\text{st}}$ and $2^{\text{nd}}$ neighbor of $a_1$ (See Fig. \ref{fig:C_2[a_1]}). Observe that, $a_1\notin N_2[a_k]$ for each $k\in \{4,5,6\}$, since $C_2[a_1]$ is the circle with largest radius. Therefore, each pair among the three pairs $\{a_1, a_4\}$, $\{a_1, a_5\}$ and $\{a_1, a_6\}$ is a subset of either $N_2[a_2]$ or $N_2[a_3]$. 
    By Pigeonhole Principle, there are at least two pairs among the three pairs $\{a_1, a_4\}$, $\{a_1, a_5\}$ and $\{a_1, a_6\}$ which are subsets of either $N_2[a_2]$ or $N_2[a_3]$, but both $N_2[a_2]$ and $N_2[a_3]$ have size $3$ which is a contradiction of $\MP(P)=1$.
    Therefore,  $\MMP(2)\leq 6$. 
    
    Furthermore, we have point sets of sizes $3,4$ and $5$ (See Fig. \ref{fig:mp2}) that have multipacking number $1$. We depict the point set $P$ in Fig. \ref{fig:mp2} forming a convex pentagon. Every point $p_i \in P$ has $p_{i-1}$ and $p_{i+1 (mod~ 5)}$ as its two nearest neighbors. As a result no two points can be present in the multipacking set of $P$. The correctness of the figure can be followed from the fact that each circle (marked with different colours) containing a point $p_i \in P$ as center and the second nearest neighbor of $p_i$ on its boundary, has only one other point (namely the first nearest neighbor of $p_i$) in the interior. This example shows $\MMP(2) > 5$ implying, $\MMP(2)= 6$.
    \end{proof}

\section{$1$-multipacking in $\mathbb{R}^2$}
\label{sec:1,1/2 Multipacking in a 2D plane} 

We note that \cref{algo:2_mp} in subsection \ref{subsec:AlgoR1} can not be adapted for a point set $P \subseteq \mathbb{R}^2$. Since we can not ensure the sorted order of the points with respect to their $x$ or $y$ coordinate, the maximum cardinality of the output can not be guaranteed. Thus it is interesting to address $r$-multipacking while $r < n-1$. 


The NNG (Nearest-Neighbor-Graph) of $P$ is a forest when considered as an undirected graph with $2$-cycles at the leaves. Then, a maximum $1$-multipacking for $P$ can be computed in polynomial time by computing the maximum independent set of NNG of $P$ \cite{eppstein1997nearest}.

We write this observation as the following lemma.

\begin{lemma}
\label{lem:1,1/2 MP 1}
    Let $P$ be a point set on the $2$D plane and the $NNG$ of $P$ be $G$. The maximum independent sets of $G$ are the maximum $1$-multipackings of $P$.
\end{lemma}

\section{$2$-multipacking in $\mathbb{R}^2$}
\label{sec:2,1/2 Multipacking in a 2D plane}
Next we study the $2$-multipacking problem  in $\mathbb{R}^2$. We are interested in studying the hardness of finding a maximum $2$-multipacking of a given set of points $P$ in the plane. 

We construct a graph $G_P=(V,E)$ from the given point set $P$. Here the set of vertices $V=P$. For every vertex $v \in V$, we introduce three edges $vu_1,vu_2,$ and $u_1u_2$  where $u_1,u_2 \in V$ are the first and second neighbors of $v$ respectively. The following lemma shows the relation between the independent set of $G_P$ and a $2$-multipacking of $P$.

\begin{lemma}
    \label{lem:graph}
    A set $M$ is a $2$-multipacking of $P$ if and only if the vertices in the graph $G_P$ corresponding to the members of $M$ form an independent set in $G_P$.
\end{lemma}
\begin{proof}
    It follows from the definition of $2$-multipacking that for every tuple $(v,u_1,$ $u_2)$, as described above, at most one can be chosen as a member of $M$. Since each edge of $G_P$ is incident on a pair of vertices appearing in a tuple, there is no edge between two members of $M$ appearing in $G_P$. Thus the vertices corresponding to the members of $M$ form an independent set of $G_P$. The converse follows from a similar argument.
\end{proof}

Thus, finding a maximum independent set of $G_P$ is equivalent to finding a maximum $2$-multipacking of $P$. In the rest of this section, we show that finding a maximum independent set of $G_P$ is \textsc{NP-hard}, and we devise approximation and parameterized solutions to this problem.

\subsection{NP-hardness}\label{subsec:np-hardness}

We show that finding a maximum independent set in the graph $G_P$ is \textsc{NP-hard} by reducing the well-known \textsc{NP-complete} problem \emph{Planar Rectilinear Monotone (PRM) 3-SAT}~\cite{de2012optimal} to this problem. This variant of the boolean formula contains three literals in each clause and all of them are either positive or negative. Moreover, a planar graph can be constructed for the formula such that a set of axis parallel rectangles represent the variables and the clauses with the following properties. The rectangles representing the variables lie on the $x$-axis. Whereas rectangles representing the clauses that contain the positive (negative) literals lie above (below) the $x$-axis. Additionally, the rectangles for the clauses can be connected with vertical lines to the rectangles for the variables that appear in the clauses.

Given a PRM 3-SAT formula $\phi$ with $t(>1)$ variables and $m(>1)$ clauses, we construct a planar point set $P_\phi$ such that $\phi$ is satisfiable if and only if $G_{P_\phi}$ has a maximum independent set of a given size. First, we construct a point set for the \emph{variable gadget} for each variable that has only two subsets as the maximum independent set in $G_{P_\phi}$. Then we construct another point set as clause gadgets for each clause in $\phi$ such that the clause gadget can incorporate only one vertex in the independent set only if at least one variable gadget has the desired set chosen as the maximum independent set. The clause and variable gadgets are connected via \emph{connection gadgets} which are two point sets referred to as \emph{translation} and \emph{rotation gadgets}. We describe the construction of the point sets formally below. 


\begin{figure}
    \centering
    \includegraphics[page=1,width=0.6\textwidth]{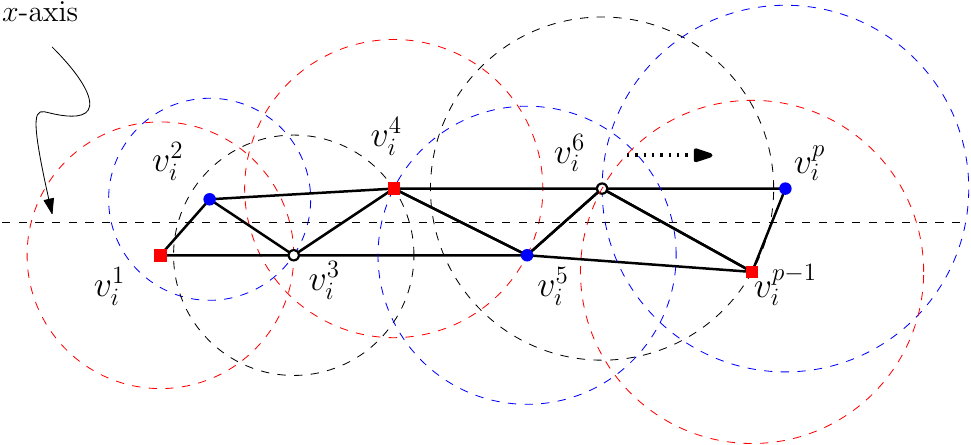}
    \caption{A variable gadget in $G_{P_\phi}$.}
    \label{fig:variable}
\end{figure}

\begin{figure}
\centering
    \includegraphics[page=2,width=0.75\textwidth]{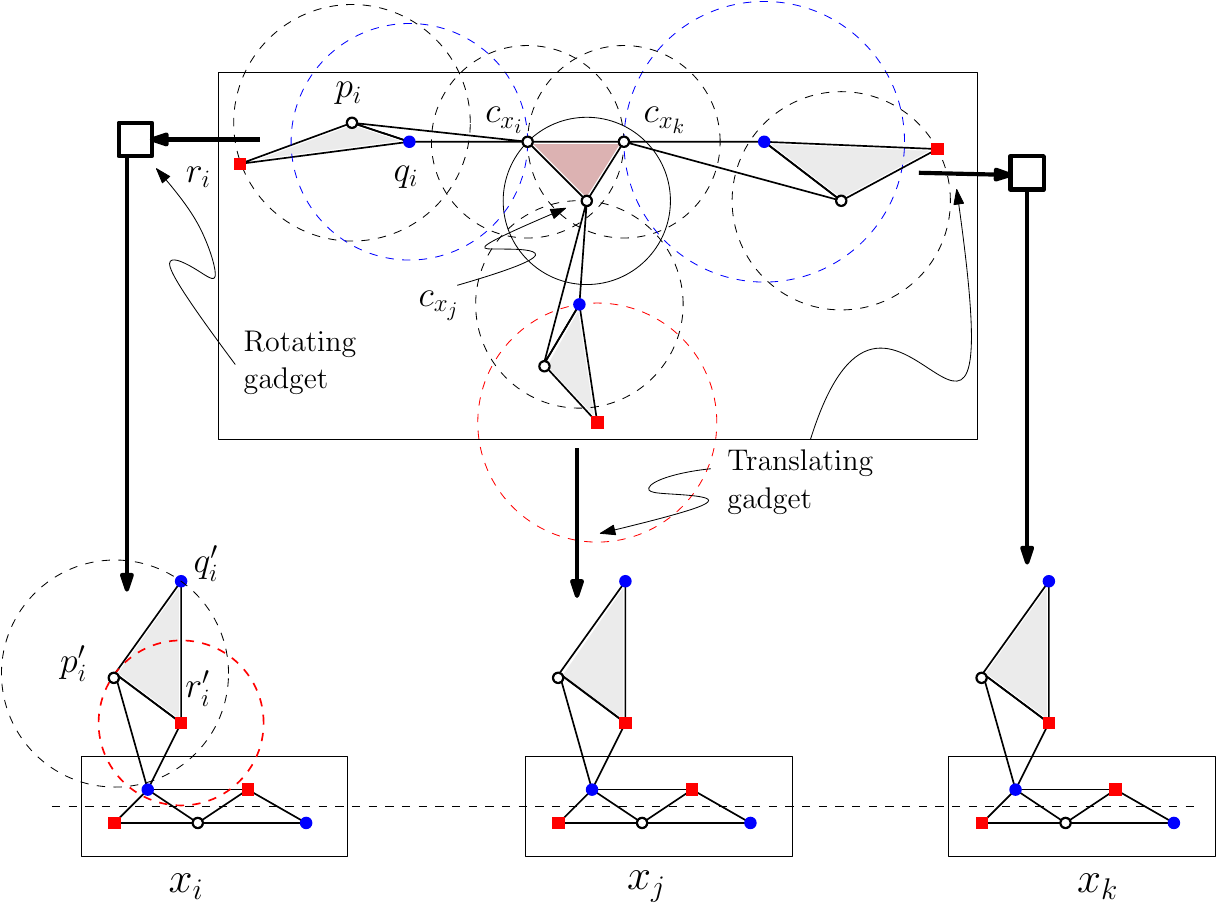}
    \caption{A clause gadget and its connection with the variable gadgets.}
    \label{fig:clause}
\end{figure}

\vspace{0.3cm}
{\bf Variable gadget.} We construct a point set for each variable $x_i$ in $\phi$ consisting of $p=6(m-1)+2$ points.  
A variable gadget is depicted in Figure~\ref{fig:variable} with $p=8$. Each point is considered as the vertex in the graph $G_{P_\phi}$. The leftmost point is $v_i^1$ having $v_i^2$ and $v_i^3$ as its first and second neighbor respectively. Thus $v_i^1$ introduces three edges $v_i^1v_i^2$, $v_i^1v_i^3$, and $v_i^2v_i^3$ in $G_{P_\phi}$. The next point from the right is $v_i^2$ which has $v_i^1$ and $v_i^3$ has its first and second neighbor. Since the edges corresponding to the adjacency of $v_i^2$ are already drawn, it does not introduce any new edges. Then the third point $v_i^3$ introduces two new edges namely $v_i^3v_i^4$ and $v_i^2v_i^4$. Continuing in a similar fashion $v_i^{p-1}$ introduces two edges $v_i^{p-1}v_i^{p}$ and $v_i^6v_i^p$. Finally, $v_i^{p}$ does not introduce any new edges having $v_i^{p-1}$ and $v_i^6$ as its first and second neighbors respectively. The second neighborhoods of the points are indicated with colored (referring to the online version in case of gray-scale documents) circles in the figure. The leftmost red circle contains the first and second neighbor of $v_i^1$ in Figure~\ref{fig:variable}. 
The following observation holds from the structure of the variable gadget in $G_{P_\phi}$. 

\begin{observation}
    \label{obs:variable}
    The maximum independent set in a variable gadget with $6(m-1)+2$ vertices, as described above, is of size $2m-1$. There are exactly two options for selecting these two sets, either the sets consisting of the vertices $V_i^T=\{v_i^1,v_i^4,\dots,v_i^{p-1}\}$ or the sets consisting of the vertices $V_i^F=\{v_i^2,v_i^5,\dots,v_i^{p}\}$. Moreover, we can place the variable gadgets in such a way that the vertices $v_i^2,v_i^4,\dots,v_i^{p-1}$ lie above the $x$-axis and the other vertices lie below the $x$-axis.
\end{observation}

We interpret the assignments of the variable $x_i$ to be true (false resp.) if  $V_i^T$ ($V_i^F$) is chosen as the maximum independent set in the variable gadget corresponding to $x_i$. Now we describe the structure of a clause gadget. 

\vspace{0.3cm}
{\bf Clause gadget.} A clause gadget consists of a point set such that the corresponding graph contains $7$ triangles. We describe the construction for a positive clause $C=(x_i \lor x_j \lor x_k)$ and the construction is analogous for the other clauses. The \emph{central} triangle of the gadget in $G_{P_\phi}$ consists of three vertices $c_{x_i},c_{x_j}$, and $c_{x_k}$ as shown (with pink shade) in Figure~\ref{fig:clause}. The triangle is introduced by placing $c_{x_j}$ and $c_{x_k}$ as the first and second neighbor of $c_{x_i}$ respectively. 
Then we draw three other vertices for each variable appearing in the clause. We describe the position of them for $x_i$ and the others are analogous. $p_i,q_i$ and $r_i$ are the three vertices placed \emph{near} $c_{x_i}$ such that $q_i$ has $c_{x_i}$ as its first neighbor and $p_i$ as its second neighbor. This introduces $\triangle q_ip_ic_{x_i}$ in $G_{P_\phi}$. Then $p_i$ has $q_i$ and $r_i$ as its first and second neighbor respectively introducing $\triangle p_iq_ir_i$ in $G_{P_\phi}$. We adjust the neighborhood of the other vertices are adjusted in such a way that no other triangle is introduced in $G_{P_\phi}$. For example, the first and second neighbors of $r_i$ are $q_i$ and $p_i$ respectively and the first and second neighbors of $c_{x_j}$ are $c_{x_i}$ and $c_{x_k}$ respectively. Considering the position of the points in the clause gadget following observation holds for the maximum independent set in the induced graph with the points in the clause gadget as the vertices. 

\begin{observation}
    \label{obs:clause}
    The maximum independent set in a clause gadget as described above has size $4$ if and only if at least one of the three vertices $r_i,r_j$, and $r_k$ is chosen in the independent set.
\end{observation}

 We consider an ordering of the clauses from left to right in the planar embedding of $\phi$. Let $C$ be the $\xi$-th clause that is connected with the variable $x_i$ from left to right. We \emph{translate} each of the triangles $\triangle p_iq_ir_i$, $\triangle p_jq_jr_j$ and $\triangle p_kq_kr_k$ \emph{near} to the corresponding variable gadgets to \emph{connect} them. We place three points $p'_i,q'_i$,and $r'_i$ near the point $v_i^{2+6(\xi-1)}$ such that $r'_i$ has $v_i^{2+6(\xi-1)}$ and $p'_i$ as its first and second neighbour. We call the vertex $v_i^{2+6(\xi-1)}$ as the connecting vertex of $C$ and $x_j$. Then we construct a point set such that the maximum independent set induced by these points as vertices in $G_{P_\phi}$ has the following property. $p_i$($q_i$ or $r_i$ resp.) can be chosen in the maximum independent set if and only if $p'_i$ ($q'_i$ or $r'_i$) is also chosen in the maximum independent set. We call this property as the \emph{connection} of the clause to the variable gadgets. Once the connections to every clause and its variables are complete we have the following observation.

\begin{observation}
    \label{obs:assign}
    The vertex $c_{x_i}$ can be chosen in an independent set in $G_{P_\phi}$ only if the maximum independent set of the variable gadget corresponding to $x_i$ is chosen to be $V_i^T$.
\end{observation}

\begin{figure}[t]
    \centering
    \includegraphics[page=3,width=.5\textwidth]{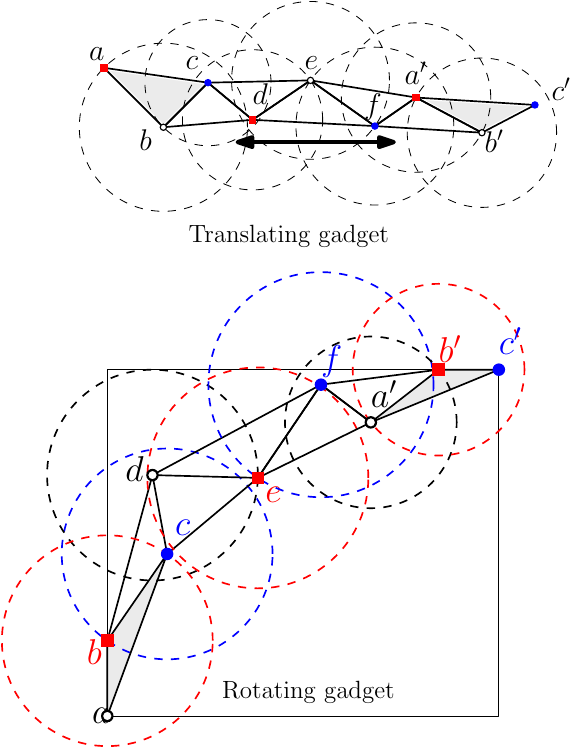}
    \caption{Translation and rotation of a triangle in $G_{P_\phi}$.}
    \label{fig:rt}
\end{figure}

The remaining work is to achieve the connection from $C$ to $x_i,x_j$, and $x_k$. We use two point sets (gadgets) to perform this move. The first one is a \emph{translating gadget} and the second one is a \emph{rotating gadget} depicted in Figure~\ref{fig:rt}. The following observation can be verified from the structure of the gadgets.

\begin{observation}
    \label{obs:translate}
    Both the gadgets contain $6$ points inside them and the size of the maximum independent set is $2$.
\end{observation}

 The translating gadget starts with a copy of a triangle and \emph{propagates} the choice of the vertex chosen in an independent set in the underlying graph. The rotating gadget does the same thing but the final triangle in this gadget is a copy of the initial triangle in the gadget but rotated at a right angle. Thus we can use mirror-reflections of the rotating gadget depicted in the figure to achieve the rotating angle to be $180^{\circ}$ and $270^{\circ}$. Together we call them the connecting gadgets.  

 \begin{figure}[t]
    \centering
    \includegraphics[page=4,width=.7\textwidth]{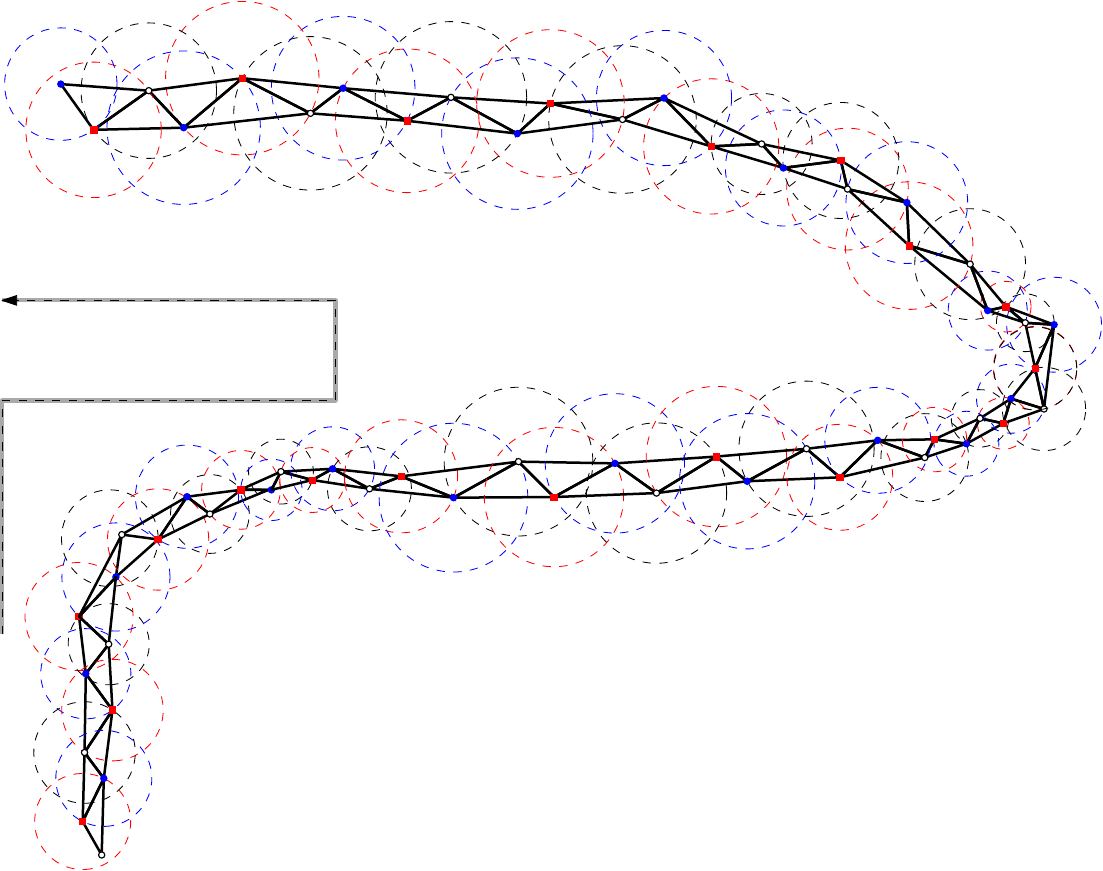}
    \caption{An example of embedding of the connecting gadgets.}
    \label{fig:emb}
\end{figure}


Now the following observation is evident from the structure of the clause gadget and its connection with the variables.

\begin{observation}
    \label{obs:number}
    Let $C=(x_i \lor x_j \lor x_k)$ one clause that is connected with its variables in $G_{P_\phi}$ with $\nu$ connecting gadgets then the maximum independent set for the point set has size $3(2m+1)+(\nu+4)$ if and only if at least one of the three variables is assigned to be true.  
\end{observation}

Now we adjust our drawing of the whole point set for $G_{P_\phi}$ such that the size of the independent set is a fixed number depending on $\phi$.  We place each clause gadget $C=(x_i \lor x_j \lor x_k)$ in such a way that the connecting gadgets for $x_j$ do not contain a rotating gadget for the connection of $c_{x_j}$ to its connecting vertex in $x_j$. We consider the fact that the nearest clause gadget from the $x$-axis can be connected with its variables by two rotating gadgets for the first and third variables and one translating gadget for the middle variable appearing in the clause.  
Then we compute the maximum number of connecting gadgets required to connect a clause with the variables.
Let $l_v$ and $w_v$ denote the length and width of the rectangles representing the variables in the planar drawing of $\phi$. We place our variable gadgets inside the rectangles. So $l_v=O(m)$ as we have placed $p=6(m-1)+2$ points in each variable gadget. Since there are $t$ variables and every clause rectangle can be connected with the variable rectangles using vertical lines, then the maximum length of a rectangle representing a clause could be $tl_v+\varepsilon$ for some small constant $\varepsilon$. On the other hand, if $w_c$ and $h_c^0$ denote the width of the clause-rectangle and the height (distance from $x$-axis) of the nearest clause-rectangle from the $x$-axis, then the distance of the farthest clause rectangle from the $x$-axis is $mw_c+h_c^0$. 
Then the maximum number of connecting gadgets required for connecting a clause gadget to its variables is at most  $3mw_c+2tl_v$.
Thus the clause gadget together with the translating and rotating gadget consists of $6(3mw_c+2tl_v)+12$ points. And thus contributing a $2(3mw_c+2tl_v)+4$ size maximum independent set. We draw our gadget in such a way that $w_c$ contains $6$ and $l_v$ contains $p$ points. Thus, every clause contains $6(18m+2tp)+12$ points together with its connecting gadgets. 

\begin{figure}[ht]
\centering
    \includegraphics[width=.65\textwidth,page=5]{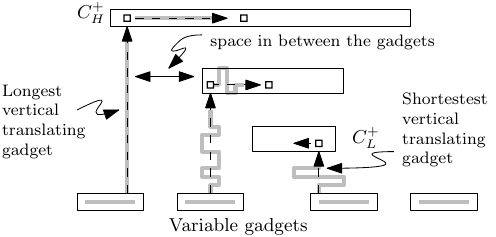}
    \caption{Convolution of the connecting gadgets in $G_{P_\phi}$.}
    \label{fig:convolution}
%
\end{figure}

To achieve this drawing, We expand the planar embedding of $\phi$ by a scalar $\alpha = (10tm)^2$. The size of the gadgets remains the same but the space between them increases such that there is a distance of at least $\alpha$ between any two connecting gadgets to accommodate the total number of points. The \emph{convolution} to accommodate the extra points for the connection is depicted in Figure~\ref{fig:convolution}. A lower-level diagram for the embedding of such convolution with translating and rotating gadgets is depicted in Figure~\ref{fig:emb}. 

Then we claim the following lemma.

\begin{lemma}
    \label{lem:NP}
    Given a PRM-3SAT formula $\phi$ we can construct a planar point set $P_\phi$ such that $G_{P_\phi}$ has a maximum independent set of size $t(m+1)+m(2(18m+2tp)+4)$ if and only if $\phi$ is satisfiable.
\end{lemma}

\begin{proof}
    If $\phi$ is satisfiable then every clause contains at least one variable that has been assigned as true. We choose $m+1$ vertices as independent set in each variable gadgets according to the truth assignment. Now for every clause $C=(x_i \lor x_j \lor x_k)$, we choose one vertex from $c_{x_i},c_{x_j}$, and $c_{x_k}$ whichever achieves a true value in the assignment. Then we choose $2(18m+2tp)$ vertices from the translation gadgets into the independent set. Since the joining vertex of  the variable gadget of a true variable will not be present in the independent set, it will not conflict with the vertex chosen from $\triangle c_{x_i}c_{x_j}c_{x_k}$. Thus we can choose $t(m+1)$ vertices from the variable gadgets and $2(18m+2tp)+4m$ vertices from the clause gadgets in the independent set. 

    On the other hand, the translation gadgets have three choices for choosing an independent set of size $2(mw_c+3tl_v)$ and the clause gadgets can contribute one vertex each towards the independent set only if at least one vertex can be chosen from the central triangle $\triangle c_{x_i}c_{x_j}c_{x_k}$. Thus we must choose the remaining $t(m+1)$ vertices in the independent set from the $t$ variable gadgets. Thus we can ensure if all clause gadgets together with their translation gadgets contributes $2(18m+2tp)+4$ vertices in the independent set then the corresponding assignment satisfies $\phi$.  
\end{proof}




Since it is possible to check the correctness of a solution for an independent set we can conclude the following theorem. 


\NPC*




\subsection{An Approximation and a Parameterised Algorithm}\label{subsec:approximation-algorithm}

In this subsection, we proved that the maximum degree of $G_P$ is bounded.

\begin{lemma}\label{lem:degree_bounded_by_17}
    The maximum degree of the graph $G_P$ is at most $17$.
\end{lemma}

\begin{figure}[t]
    \centering
    \includegraphics[width=0.45\textwidth]{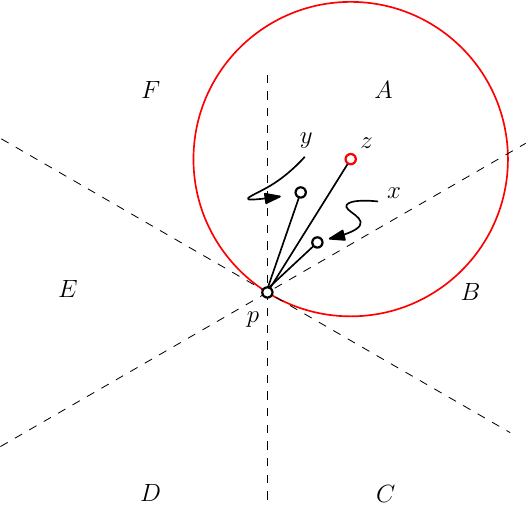}
    \caption{The maximum degree of $G_P$ is at most 17.}
    \label{fig:degree17}    
\end{figure}

\begin{proof}
Let $p$ be any point of $P$. Consider the graph $G_P=(V,E)$ that we defined in the beginning of the Section \ref{sec:2,1/2 Multipacking in a 2D plane}. Here $V=P$.  We want to show that the degree of $p$ in $G_P$ is bounded by $17$.  Recall that, $N_r[v]$ is the subset of $P$ that includes the $r$ nearest points of $v$ and the point $v$ itself. We denote the $r$-th neighbour of $v$ by $n_r(v)$.  Here we define some set of points of $P$ that will help us to prove this lemma.
Let $S_1=\{v\in P:n_1(v)=p \text{ or }n_2(v)=p\}$,  $S_2=\{n_1(p),n_2(p)\}$ and $S_3=\{v\in P\setminus S_1\cup S_2:p,v\in N_2[w] \text{ for some }w \in P\}$. Note that, the degree of $p$ in $G_P$ is $|S_1\cup S_2\cup S_3|$. Therefore, we have to show that $|S_1\cup S_2\cup S_3|\leq 17$.



\vspace{0.3cm}
\noindent
\textbf{Claim \ref{lem:degree_bounded_by_17}.1. }  $|S_1\cup S_2|\leq 12$. 
\begin{claimproof} We partition the plane by three concurrent lines meeting at $p$ that generate $6$ wedges of equal angles (See Fig.\ref{fig:degree17}). Let $A,B,C,D,E$ and $F$ be the $6$ wedges. Our goal is to prove that no wedge can contain more than two points from the set $S_1\cup S_2$. We prove this by contradiction. Without loss of generality assume that the wedge $A$ contains at least $3$ points from the set $S_1\cup S_2$. Let $S_A$ be the set of all points which belong to both $S_1\cup S_2$ and $A$. Let $z\in S_A$ be the farthest point from $p$ among all the points from $S_A$. Therefore $z\notin S_2$. This implies $p\in N_2[z]$. Let $x,y\in S_A\setminus\{z\}$. Now consider the triangle $\triangle xpz$. This is not an equilateral triangle since $px<pz$. Moreover, the angle $\angle xpz\leq 60^{\circ}$. This implies either $\angle pxz>60^{\circ}$ or $\angle pzx>60^{\circ}$.  Therefore $xz$ is not the largest side of $\triangle xpz$. Now $px<pz$ implies $pz$ is the largest side of $\triangle xpz$. Therefore $x\in N_2[z]$. Similarly, we can show that $y\in N_2[z]$. Therefore, $x,y,z,p\in N_2[z]$. This is a contradiction.   
\end{claimproof}

\vspace{0.3cm}
\noindent
\textbf{Claim \ref{lem:degree_bounded_by_17}.2. }  $|S_3|\leq 5$. 
\begin{claimproof} Suppose $|S_3|\geq 6$. Define a function $f:S_3\rightarrow P$ where we say $f(v)=w$ if either $v$ is the first neighbor of $w$ and $p$ is the second neighbor of $w$ or $v$ is the second neighbor of $w$ and $p$ is the first neighbor of $w$. Note that $f$ is an injective function. Therefore, the size of the image set, $|f(S_3)|\geq 6$. Moreover, $f(S_3)\subset S_1\cup S_2$. Therefore, $f(S_3)\cap S_3=\phi$. Now we remove all the points of $S_3$ from $P$. Let $P'=P\setminus S_3$. In $P'$, the first neighbor of each point of $f(S_3)$ is $p$. Therefore, there are at least $6$ points in $P'$ that have $p$ as their first neighbor. Then there exist two points $v_1,v_2\in f(S_3)$ such that $\angle v_1pv_2\leq 60^{\circ}$. We have $pv_1\neq pv_2$, since each vertex has only one $r$-th neighbor, for any $r$. Without loss of generality assume that $pv_1> pv_2$. This implies $pv_1$ is the largest side in $\triangle v_1pv_2$. Therefore, $p$ cannot be the first neighbor of $v_1$. This is a contradiction. 
\end{claimproof}

From the above two claims, we can say that $|S_1\cup S_2\cup S_3|\leq 17$.  
\end{proof}

Using this we provide approximation and parameterized solutions to the $2$-multipacking problem of a given set of points $P$ in the plane. We use existing techniques for computing the maximum independent set for bounded degree graphs.

The problem of finding the Maximum Independent Set in a bounded degree graph is denoted by \textit{MAX IS-$\Delta$} where the maximum node degree of the graph is bounded above by $\Delta$. The following result states the approximation bound of the \textit{MAX IS-$\Delta$} problem.

\begin{theorem}[Berman and  Fujito \cite{berman1999approximation}]\label{thm:App17}
    \textit{MAX IS-$\Delta$} can be approximated in polynomial time within a ratio arbitrarily close to $(\Delta + 3)/5$ for all $\Delta \geq 2$.
\end{theorem}

Theorem \ref{thm:App17} and Lemma \ref{lem:degree_bounded_by_17} yield the following.

\begin{restatable}{theorem}{thmapprox}\label{thm:approx}
 A maximum $2$-multipacking of a given set of points $P$ in $\mathbb{R}^2$ can be approximated in polynomial time within a ratio arbitrarily close to $4$.
\end{restatable}


In 2019, Bonnet et al. have shown the following:

\begin{theorem}[Bonnet et al.\cite{bonnet2019maximum}]\label{thm:FPT_for_bounded_degree_graph}
    For graphs of $n$ vertices with degree bounded by $\Delta$, we can compute an independent set of size $k$ in  time $(\Delta+1)^kn^{O(1)}$, if it exists.
\end{theorem}

Lemma \ref{lem:degree_bounded_by_17} and Theorem \ref{thm:FPT_for_bounded_degree_graph} yield the following result.

\begin{restatable}{theorem}{thmFPT}\label{thm:FPT}
 A $2$-multipacking of size $k$ of a given set of $n$ points in $\mathbb{R}^2$ can be computed in time $18^kn^{O(1)}$, if it exists.
\end{restatable}

\section{Conclusion}
\label{sec:conclusion}
This paper studies some multipacking problems in a geometric setting under the Euclidean distance metric.  An intriguing open topic in this area is addressing the difficulty of computing a maximum multipacking ($(n-1)$-multipacking) in general. We know neither the algorithmic difficulties nor the limitations as of yet for the $(n-1)$-multipacking problem.

In this paper, we have given the value of $\MMP(2)$. A natural direction for future research is determining the exact values of $\MMP(t)$ for $t \geq 3$.

Additionally, the definition of $r$-multipacking in geometry, inspired by obnoxious facility location and graph-theoretic multipacking, can be generalized by introducing a fractional parameter $k \in [0,1]$. Specifically, the condition on a set $M$ could be generalized to $|N_s[v] \cap M| \leq k(s+1)$ for each $v \in P$. Varying $k$ may lead to different bounds on the multipacking number and introduce new computational challenges.

Further extensions of this work could explore multipackings in higher-dimensional Euclidean spaces $\mathbb{R}^d$, where geometric constraints and packing densities may yield richer structural and algorithmic results.





\bibliographystyle{elsarticle-num}
\bibliography{ref.bib}






\end{document}